\documentclass[a4paper,12pt]{article}
\usepackage{latexsym,amssymb}

\newcommand{\be}{\begin{equation}}
\newcommand{\ee}{\end{equation}}
\newcommand{\bea}{\begin{eqnarray}}
\newcommand{\eea}{\end{eqnarray}}
\newcommand{\ba}{\begin{array}}
\newcommand{\ea}{\end{array}}

\newcommand{\ds}{\displaystyle}

\newcommand{\BPSA}{\emph{BPS Ansatz}\,}
\newcommand{\diag}{{\tt diag}}

\makeatletter \@addtoreset{equation}{section} \makeatother

\begin{document}
\title{\bf Attractors with Vanishing Central Charge}
\date{}
\author{S.~Bellucci${\,}^a$, A.~Marrani${\,}^{a\;b}$, E.~Orazi${\,}^c$
and~A.~Shcherbakov${\,}^a$\footnote{On leave of absence from JINR,
Dubna, Russia}} \maketitle
\begin{center}
{${}^a$ \it INFN - Laboratori Nazionali di Frascati, \\
Via Enrico Fermi 40, 00044 Frascati, Italy\\
\texttt{bellucci@lnf.infn.it, marrani@lnf.infn.it,
ashcherb@lnf.infn.it}}

\vspace{15pt}

{${}^b$ \it Museo Storico della Fisica e\\
Centro Studi e Ricerche ``Enrico Fermi''\\
Via Panisperna 89A, 00184 Roma, Italy}

\vspace{15pt}

{${}^c$ \it Dipartimento di Fisica,Politecnico di Torino,\\
Corso Duca degli Abruzzi 24, 10129 Torino, Italy\\
and INFN - Sezione di Torino, Italy\\
\texttt{emanuele.orazi@polito.it}}
\end{center}

     \thispagestyle{empty}



\begin{abstract}
We consider the Attractor Equations of particular
$\mathcal{N}=2$,~$d=4$ supergravity models whose vector
multiplets' scalar manifold is endowed with homogeneous symmetric
cubic special K\"{a}hler geometry, namely of the
so-called~$st^{2}$ and~$stu$ models. In this framework, we derive
explicit expressions for the critical moduli corresponding to
non-BPS attractors with vanishing $\mathcal{N}=2$ central charge.
Such formul\ae\ hold for a generic black hole charge
configuration, and they are obtained without formulating any
\textit{ad hoc} simplifying assumption. We find that such
attractors are related to the $\frac{1}{2}$-BPS ones by complex
conjugation of some moduli. By uplifting to~$\mathcal{N}=8$,~$d=4$
supergravity, we give an interpretation of such a relation as an
exchange of two of the four eigenvalues of the $\mathcal{N}=8$
central charge matrix~$Z_{AB}$. We also consider non-BPS
attractors with non-vanishing $\mathcal{Z}$; for peculiar charge
configurations, we derive solutions violating the Ansatz usually
formulated in literature. Finally, by group-theoretical
considerations we relate Cayley's hyperdeterminant (the invariant
of the $stu$ model) to the invariants of the $st^{2}$ and of the
so-called $t^{3}$ model.
\end{abstract}


\section{Introduction}

The Attractor Mechanism in extremal black holes (BHs)~\cite
{FKS,Strom,FK1,FK2,FGK} has recently been investigated in
depth, and various advances along such a line of research has been performed~\cite{Sen-old1}--
\cite{CFM1}.

The horizon geometry of extremal black holes in $d=4$ space-time
dimensions is the direct product of two spaces with non-vanishing,
constant (and opposite) curvature, namely it is the
Bertotti-Robinson\ geometry~\cite {BR1,BR2,BR3}:
\begin{equation}
AdS_{2}\times S^{2}=\frac{SO(1,2)}{SO(1,1)}\times
\frac{SO(3)}{SO(2)}. \label{BR-geometry}
\end{equation}
In the framework of $\mathcal{N}=2$, $d=4$ supergravity, such an
horizon geometry is associated to the maximal $\mathcal{N}=2$
supersymmetry algebra~$\mathfrak{psu}(1,1\left|2\right)$, which is
an interesting example of superalgebra containing not Poincar\'{e}
nor semisimple groups, but direct products of simple groups as
maximal bosonic subalgebra. Indeed, in this case the maximal
bosonic subalgebra is $\mathfrak{so}(1,2)\oplus \mathfrak{su}(2)$
(with related maximal spin bosonic subalgebra
$\mathfrak{su}(1,1)\oplus \mathfrak{su}(2)$), matching the
corresponding bosonic isometry group of the Bertotti-Robinson
metric (\ref{BR-geometry}).

In this context, the attractor configurations of the scalars at
the event horizon of the BH have been recognized to fall into
three distinct classes (\cite{BFM,BFGM1,AoB-book}; see
also~\cite{ADFT} for a recent review):
\begin{enumerate}
\item the $\frac{1}{2}$-BPS class, known since~\cite {FKS,Strom,FK1,FK2,FGK}, preserving four supersymmetries out of the
eight pertaining to the asymptotical Minkowski space-time
background;

\item the non-BPS class with non-vanishing $\mathcal{N}=2$ central charge at the horizon, which does not preserve any
supersymmetry at all;

\item the non-BPS class with vanishing $\mathcal{N}=2$ central charge at the horizon; thus, for this class the complete
breakdown of supersymmetry at the BH event horizon is associated
with the lack of central extension of $\mathfrak{psu}(1,1\left|
2\right) $.\medskip
\end{enumerate}

In this work we address the issue of the explicit determination of
the non-BPS scalar configuration with vanishing $\mathcal{N}=2$
central charge, in the framework of peculiar $\mathcal{N}=2$,
$d=4$ supergravities coupled to $n_{V}=2$ and $3$ vector
multiplets, namely for the so-called $st^{2}$ and $stu$
models~\cite{BFGM1}.

Such models belong to the broad class of the
so-called~$d$-geometries (\cite {dWVVP}; see Sect.~\ref{SKG} for
elucidation), whose explicit~$\frac{1}{2}$-BPS attractors, for
generic BH charges and generic~$n_{V}$, are known
after~\cite{Shmakova} (the~$stu$ model was previously investigated
in~\cite{BKRSW}). Recently, in~\cite{TT} the $d$-geometries with
generic $n_{V}$ were reconsidered, and the non-BPS attractors with
non-vanishing central charge were explicitly determined for a
peculiar choice of BH charges.

In our investigation, we find that, for a general charge
configuration, the non-supersymmetric attractors with
vanishing~$\mathcal{N}=2$ central charge always violate the Ansatz
used in~\cite{TT}. Furthermore, due to the high symmetry of the
scalar geometries analyzed, they turn out to be intimately related
to the $\frac{1}{2}$-BPS attractors.\medskip

The plan of this work is as follows.

In Sect.~\ref{SKG} we briefly recall the foundations of the
special K\"{a}hler (SK) geometry endowing the vector multiplets'
scalar manifold of~$\mathcal{N}=2$,~$d=4$ supergravity, focussing
on the so-called SK~$d$-geometries, and limiting ourselves to the
sole quantities needed in the subsequent treatment.
Sect.~\ref{st^2} is devoted to the~$st^{2}$ model, the simplest
symmetric model in which non-supersymmetric attractors with
vanishing central charge appear; the violation of the Ansatz
of~\cite{TT} is pointed out, and the explicit form of such
attractors, along with the relation with the
well-known~$\frac{1}{2}$-BPS ones, is derived. As a byproduct of
our approach, we also obtain a one-parameter family of non-BPS
attractors with non-vanishing central charge which violate the
Ansatz of~\cite{TT}, showing that it actually implies some loss of
generality in the context of SK $d$-geometries. In Sect.~\ref{stu}
we perform an analogous analysis in the $stu$
model~\cite{Duff-stu,BKRSW,K3}. We derive the explicit expression
of non-supersymmetric attractors with vanishing central charge,
and elucidate how they related with the supersymmetric ones.
Finally, in Sect.~\ref{Cayley-hyperdeterminant} we relate, by
simple group-theoretical considerations, Cayley's hyperdeterminant
to the quartic (and unique) invariants of the~$U$-duality groups
of the models $t^{3}$~\cite {Saraikin-Vafa-1} and $st^{2}$.
Concluding remarks and an outlook can be found in the final Sect.
\ref{Conclusion}.

\section{\label{SKG}Special K\"{a}hler Geometry}

In this section we briefly recall some notions of the special
K\"{a}hler (SK) geometry underlying the vector multiplets' scalar
manifold of~$\mathcal{N}=2$, $d=4$ supergravity coupled to $n_{V}$
vector multiplets. Our treatment is far from exhaustive, as we
introduce only the quantities needed in the subsequent
computations (for the notation, explanation and extensive
treatment, see \textit{e.g.}~\cite{CDF-review} and Refs. therein).

Once a holomorphic prepotential function~$F\left( X\right) $ of
the sections~$X^{\Lambda }$~($\Lambda =0$, $1$,...,$n_{V}$) is
given, one can derive all the fundamental quantities in the
framework of SK geometry. The K\"{a}hler potential and the
corresponding moduli space metric are found to be
\begin{equation}
K=-ln\left[ i\left( \bar{X}^{\Lambda }\partial _{\Lambda
}F-X^{\Lambda }\overline{\strut \partial _{\Lambda }F}\;\right)
\right] \,,~~g_{i\bar{j}}=\partial _{i}\partial _{\bar{j}}K\,,
\end{equation}
where the indices~$i$ and~$\overline{j}$ refer to the
moduli~$z^{i}$ and~${\bar
z}^{\overline{j}}$~($i,\overline{j}=1,\ldots,n_{V}$ throughout),
respectively. The covariantly holomorphic~$\mathcal{N}=2$ central
charge function
\begin{equation}
\mathcal{Z}\left( z,{\bar z},p,q\right) =e^{\frac12 K\left(
z,{\bar z} \right)}W\left( z,p,q\right) \,
\end{equation}
is given in terms of the holomorphic superpotential
\begin{equation}
W\left( z,p,q\right) =q_{\Lambda }X^{\Lambda }-p^{\Lambda
}\partial _{\Lambda }F\,, \label{prep}
\end{equation}
and it can be used to calculate the so-called BH effective
potential~\cite {FGK}
\begin{equation}
V_{BH}=e^{K}\left[ g^{i\bar{j}}\left( \mathcal{D}_{i}W\right)
\overline{{\mathcal{D}}}{_{\bar{j}}\bar{W}}+W\bar{W}\right]
=|\mathcal{Z}|^{2}+g^{i\bar{j}}\left(
\mathcal{D}_{i}\mathcal{Z}\right)
\overline{\mathcal{D}}_{\bar{j}}\overline{\mathcal{Z}}\,,
\label{VBH}
\end{equation}
where $\mathcal{D}_{i}=\partial _{i}+\frac{p}{2}\partial _{i}K$
denotes the K\"{a}hler-covariant derivative acting on an object
with holomorphic K\"{a}hler weight $p$.

The Attractor Mechanism in extremal
BHs~\cite{FKS,Strom,FK1,FK2,FGK} yields that at the BH event
horizon the moduli are stabilized in terms of the
electric~$q_{0}$,~$q_{i}$ and magnetic charges~$p^{0}$,~$p^{i}$ of
the BH as they are critical points of~$V_{BH}$, \textit{i.e.} they
are solutions of the Attractor Equations (AEs) given by the
criticality conditions of $V_{BH}$. Through the fundamental
relations characterizing the special K\"{a}hler geometry the AEs
can be cast in the following form:
\begin{equation}
\mathcal{D}_{i}V_{BH}=2\overline{\mathcal{Z}}\mathcal{D}_{i}\mathcal{Z}+iC_{ijk}g^{j\bar{l}}g^{k\bar{m}}\left(
\overline{\mathcal{D}}_{\bar{l}}\overline{\mathcal{Z}}\right)
\overline{\mathcal{D}}_{\bar{m}}\overline{\mathcal{Z}} \label{AEs}
\end{equation}
where $C_{ijk}$ is the completely symmetric, covariantly
holomorphic rank-$3$ tensor, defined as
\begin{equation}
C_{ijk}\equiv e^{K}\left( \partial _{i}X^{\Lambda }\right) \left(
\partial _{j}X^{\Sigma }\right) \left( \partial _{k}X^{\Xi
}\right) \partial _{\Xi }\partial _{\Sigma }F_{\Lambda }\left(
X\right). \label{Cijk}
\end{equation}

Starting from the general structure of the criticality
conditions~(\ref{AEs}) and assuming also the
\textit{non-degeneracy} (\textit{i.e.} ~$\left. V_{BH}\right|
_{\partial V_{BH}=0}>0$) \textit{condition}, as mentioned above
the critical points of $V_{BH}$ (\textit{i.e.} the attractors of
$\mathcal{N}=2$, $d=4$ supergravity) can be classified in three
general classes (\cite{BFM,BFGM1,AoB-book}; see also~\cite{ADFT}
for a recent review):
\begin{enumerate}
\item The \textit{supersymmetric} $\frac{1}{2}$-\textit{BPS} class, determined by the constraints
\begin{equation}
\mathcal{Z}\neq 0,\qquad \mathcal{D}_{i}\mathcal{Z}=0,\,\forall i.
\label{BPS-conds}
\end{equation}
The corresponding horizon ADM squared mass~\cite{ADM} saturates
the BPS bound~\cite{BPS}:
\begin{equation}
M_{ADM}^2=V_{BH}\rule[-0.7em]{0.2pt}{1.6em}_{\,cr}=\left|
\mathcal{Z} \right|_{cr}^{2}>0, \label{V-BPS}
\end{equation}
where the BH effective potential and central charge are evaluated
at critical points.
\item The \textit{non-BPS}~$\mathcal{Z}\neq 0$ class,
determined by the constraints
\begin{equation}
\mathcal{Z}\neq 0,\qquad \mathcal{D}_{i}\mathcal{Z}\neq 0
\quad\mbox{at least for one value of~$i$}.
\label{non-BPS-Z<>0-conds}
\end{equation}
By using the properties of SK geometry, the
non-BPS~$\mathcal{Z}\neq 0$ horizon ADM squared mass is found not
to saturate the BPS bound:
\begin{equation}
M_{ADM}^{2}=V_{BH}\rule[-0.7em]{0.2pt}{1.6em}_{\,cr}= 4\left|
\mathcal{Z}\right|^2_{cr}\label{V-non-BPS-Z<>0-1}
\end{equation}

\item The \textit{non-BPS}~$\mathcal{Z}=0$ class, determined by the
constraints
\begin{equation}
\mathcal{Z}=0,\qquad\mathcal{D}_{i}\mathcal{Z}\neq 0\quad\mbox{at
least for one value of~$i$}. \label{non-BPS-Z=0-conds}
\end{equation}
As far as $g_{i\overline{j}}$ is strictly positive-definite the
non-BPS~$\mathcal{Z}=0$ horizon ADM squared mass does not saturate
the BPS bound~\cite{K1,K2,Tom}
\begin{equation}
M_{ADM}^{2}=V_{BH}\rule[-0.7em]{0.2pt}{1.6em}_{\,cr}=
g^{i\overline{j}}\left( \partial _{i}\mathcal{Z}\right)
\overline{\partial
}_{\overline{j}}\overline{\mathcal{Z}}\,\rule[-0.7em]{0.2pt}{1.6em}_{\,cr}
> \left| \mathcal{Z}\right|_{cr}^{2}. \label{V-non-BPS}
\end{equation}
In general, the $n_{V}$ complex nonlinear equations~(\ref{AEs})
are very difficult to solve. Some simplification can be achieved
by switching off some of the magnetic or electric charges, and
eventually by formulating suitable \textit{ad hoc} assumptions.
\end{enumerate}
The treatment given in the next Sections concerns particular
examples the so-called special K\"{a}hler $d$\textit{-geometries}
~\cite{dWVVP}, based on cubic holomorphic prepotentials:
\begin{equation}
F\left( X\right)
=\frac{1}{3!}d_{ijk}\frac{X^{i}X^{j}X^{k}}{X^{0}}=\left(
X^{0}\right) ^{2}f\left( z\right) ,~~f\left( z\right) \equiv
\frac{1}{3!}d_{ijk}z^{i}z^{j}z^{k},
\end{equation}
where the special coordinates~$X^{\Lambda }=\left(
X^{0},X^{0}z^{i}\right)$ have been introduced. We will deal
with~$f\left(z\right)$, by further fixing the K\"{a}hler gauge
such that~$X^{0}\equiv 1$. Special K\"{a}hler~$d$-geometries
naturally arise in the large volume limit of Calabi-Yau
compactifications of type IIA superstrings. They include all
special K\"{a}hler coset manifolds, of which symmetric spaces are
in turn a peculiar case.

As mentioned in the Introduction,~$\frac{1}{2}$-BPS solutions to
AEs~(\ref {AEs}) for~$d$-geometries are known after~\cite{BKRSW}
and~\cite{Shmakova}. Some non-BPS~$\mathcal{Z}\neq 0$ solutions to
AEs~(\ref{AEs}) for~$d$-geometries has been explicitly obtained
for a charge configuration~$q_{i}=0$ in~\cite{TT}. Such solutions
satisfy the peculiar Ansatz
\begin{equation}
z_{cr}^{i}\left( q,p\right) = p^{i}\lambda \left(q,p\right) ,
\label{TT-Ansatz}
\end{equation}
where~$\lambda $ is a complex function of the considered set of BH
charges. While general~$\frac{1}{2}$-BPS solutions always
satisfy~(\ref{TT-Ansatz}), this is not the case for~\emph{all}
non-BPS solutions, as we will show for the~$st^{2}$ and~$stu$
models. Therefore we refer to~(\ref{TT-Ansatz}) as a ``\emph{BPS
Ansatz}''.

As it will be demonstrated below, among non-BPS~$\mathcal{Z}\neq
0$ solutions there are those obeying~\emph{BPS Ansatz} as well as
those violating it, while non-BPS solutions with vanishing central
charge never satisfy~(\ref{TT-Ansatz}).

\section{\label{st^2}$st^{2}$ Model}

In the so-called~$st^{2}$ model the scalar manifold is the
rank-$2$ homogeneous symmetric space~$\left(
\frac{SU(1,1)}{U(1)}\right)^{2}$ ($dim_{\Bbb{C}}=2$),
parameterized by the moduli~$z^{1}\equiv s$ and~$z^{2}\equiv t$
with negative imaginary parts~\cite{Gilmore}. The
corresponding~$d=5$ parent theory is based on the rank-$1$ real
special manifold $SO\left( 1,1\right) $, which in turn can be
uplifted to pure $\mathcal{N}=2$, $d=6$ supergravity. In special
coordinates the prepotential reads~$f=st^{2}$.

Let us mention that the~$st^{2}$ can be obtained e.g. by a~``$t=u$
degeneracy'' of the~$stu$ model (see Sect.~\ref{stu}), or as the
element~$n=-1$ of the reducible cubic sequence of homogeneous
symmetric SK manifolds~$\frac{SU(1,1)}{U(1)}\otimes
\frac{SO(2,2+n)}{SO(2)\otimes SO(2+n)}$ (see e.g.~\cite{BFGM1} and
Refs. therein), or also as~$Solv_{SK}\left( -1,0\right) $,
\textit{i.e.} as the homogeneous symmetric~$P=0$ element of the
homogeneous (non-symmetric, except for~$P=0$)
sequence~$Solv_{SK}\left( -1,P\right)$~\cite{Tits-Satake}. As
the~$stu$ model, it is contained in all homogeneous (not
necessarily symmetric) SK~$d$-geometries.

For a given set of BH charges~$\left(
q_{0},q_{1},q_{2},p^{0},p^{1},p^{2}\right) $ the superpotential,
BH potential, contravariant metric and~$C$-tensor are respectively
given by :
\begin{equation}
\begin{array}{l}
\ds W(s,t)=q_{0}+q_{1}s+q_{2}t+p^{0}st^{2}-p^{1}t^{2}-2p^{2}st\,, \\
\\
\ds V_{BH}=\frac{i}{2\left( s-\bar{s}\right) \left(
t-\bar{t}\right)^{2}} \{ \left[ |W(s,t)| + |W(s,\bar t)| +(p^1-p^0
s)(t-\bar t)^2\right]\cdot\\
\\
\quad\quad \cdot\left[ W(\bar s,t) + W(\bar s,\bar t)
+(p^1-p^0 \bar s)(t-\bar t)^2\right]+2(|W(s,t)|^2+|W(s,\bar t)|^2)\}\\
\\
\ds g^{i\bar{j}}=-\diag\left(
(s-\bar{s})^{2},\frac{1}{2}(t-\bar{t})^{2}\right), \qquad
C_{stt}=\frac{2i}{(s-\bar{s})(t-\bar{t})^{2}}.
\end{array}
\label{st^2-geometry}
\end{equation}

Correspondingly, the two complex AEs have the form
\begin{equation}
\left\{
\begin{array}{l}
\left[ W(\bar s,t) + W(\bar s,\bar t) +(p^1-p^0 \bar s)(t-\bar
t)^2\right]^2+4W(\bar s,t)W(\bar s,\bar t) =0; \\
\\
W(\bar s,\bar t)\left[ W(s,t) + W(s,\bar t) +(p^1-p^0 s)(t-\bar
t)^2\right]+ \\
\\
+W(s,\bar t)\left[ W(\bar s,t) +W(\bar s,\bar t) +(p^1-p^0 \bar
s)(t-\bar t)^2\right]=0.
\end{array}
\right. \label{st^2-AEs}
\end{equation}

\subsection{Magnetic charge configuration}
By putting~$p^{0}=0$ and~$q_{1}=q_{2}=0$, the solutions of AEs
(\ref{st^2-AEs}) can be found to be ($s\equiv s_{1}+is_{2}$
and~$t\equiv t_{1}+it_{2}$):
\begin{eqnarray}
&&\left\{ s_{1}=t_{1}=0,\qquad s_{2}=\pm \frac{\sqrt{\strut
q_{0}p^{1}}}{p^{2}},\qquad t_{2}=-\sqrt{\strut
\frac{q_{0}}{p^{1}}}\right\}, \quad q_{0}p^{1}>0,~p^{2}\lessgtr 0; \label{W0} \\
&&\left\{ s_{1}=t_{1}=0,\qquad s_{2}=\pm \frac{\sqrt{\strut
-q_{0}p^{1}}}{p^{2}},\qquad t_{2}=-\sqrt{\strut
-\frac{q_{0}}{p^{1}}}\right\},\quad q_{0}p^{1}<0,~p^{2}\lessgtr 0.
\label{Wnon0}
\end{eqnarray}

Some of these solutions are known since~\cite{Shmakova}
($\frac{1}{2}$-BPS) and~\cite{TT} (non-BPS~$\mathcal{Z}\neq 0$
satisfying the Ansatz (\ref {TT-Ansatz})).

The analysis of the first set of solutions (\ref{W0}) yields the
following results:
\begin{center}
\begin{tabular}{l|c|c|c|c|}
\hline \rule[-0.5em]{0pt}{2em} & sign chosen for~$s_{2}$ & class &
\BPSA satisfied & BH charge domain \\ \hline
$1~$ & $+$ & non-BPS $\mathcal{Z}=0$ & no & $q_{0}p^{1}>0,\quad p^{2}<0$ \\
\hline $2~$ & $-$ & $\frac{1}{2}$-BPS & yes &
$\rule[-1.5em]{0pt}{3.5em}q_{0}p^{1}>0,\quad p^{2}>0$ \\ \hline
\end{tabular}
\end{center}
Such solutions are stable, since the corresponding critical
Hessian matrix has all strictly positive eigenvalues. The
non-BPS~$\mathcal{Z}=0$ solution~$2$ turns out to violate the
\BPSA~(\ref{TT-Ansatz}) by a sign.

On the other hand, the analysis of the possible combinations of
signs within the second set of solutions~(\ref{Wnon0}) yield the
following results:
\begin{center}
\begin{tabular}{l|c|c|c|c|}
\hline \rule[-0.5em]{0pt}{2em} & sign chosen for $s_{2}$ & class &
\BPSA & BH charge domain \\ \hline $3~~$ & $-$ & non-BPS
$\mathcal{Z}\neq 0$ & yes &
$\rule[-1.5em]{0pt}{3.5em}q_{0}p^{1}<0,\quad p^{2}>0$ \\ \hline
$4~~$ & $+$ & non-BPS $\mathcal{Z}\neq 0$ & no
&~$\rule[-1.5em]{0pt}{3.5em}q_{0}p^{1}<0,\quad p^{2}<0$ \\ \hline
\end{tabular}
\end{center}
Such solutions belong to the class non-BPS~$\mathcal{Z}\neq 0$. In
agreement with~\cite{TT}, the corresponding real form of
the~$4\times 4$ Hessian matrix has three strictly positive and one
vanishing eigenvalue. As found in~\cite{Ferrara-Marrani-2}, such a
massless Hessian mode is actually a ``flat'' direction
of~$V_{BH}$, spanning the non-BPS~$\mathcal{Z}\neq 0$ moduli
space~$SO\left( 1,1\right) $, which is the real special manifold
($dim_{\Bbb{R}}=1$) of the corresponding~$d=5$ supergravity theory
(see Table 2 of~\cite{Ferrara-Marrani-2}). Thus, solutions~$3$
and~$4$ are stable, up to a ``flat'' direction to all orders.
Solution~$4$ violates \BPSA by a sign, and thus it shows that such
an Ansatz implies some loss of generality, also when considering
only the class of non-BPS~$\mathcal{Z}\neq 0$ critical points
of~$V_{BH}$ in SK~$d$-geometries within peculiar BH charge
configurations.

Beside the sets (\ref{W0}) and (\ref{Wnon0}), the only other set
of solutions of AEs (\ref{st^2-AEs}) is given by the following
non-BPS~$\mathcal{Z}\neq 0$ family, parameterized by~$t_{1}\equiv
{\tt Re}\, t$:
\begin{equation}
s_{1}=-\frac{2q_{0}p^{1}t_{1}}{p^{2}\left(
q_{0}-p^{1}t_{1}^{2}\right) },\quad s_{2}=
-\rule[-1em]{0.05em}{2.4em}\,\frac{q_{0}+p^{1}t_{1}^{2}}{q_{0}-p^{1}t_{1}^{2}}\,
\rule[-1em]{0.05em}{2.4em}\,\sqrt{-\frac{q_{0}p^{1}}{\left(p^2
\right)^2}},\quad t_{2}=-\sqrt{-\frac{q_{0}}{p^{1}}-t_{1}^{2}}
\label{long}
\end{equation}
and defined for~$q_{0}p^{1}<0$,~$t_{1}^{2}<-q_{0}/p^{1}$. Let us
also notice that in the limit~$t_{1}\rightarrow 0$ the
solutions~(\ref{long}) coincide with the~(\ref{Wnon0}).

For all sets (\ref{W0}), (\ref{Wnon0}) and (\ref{long}) the
minimal value of the BH effective potential is
\begin{equation}
V_{BH}\rule[-0.7em]{0.2pt}{1.6em}_{\,cr}=\frac{S_{BH}}{\pi
}=2\sqrt{\left| q_{0}p^{1}\right| (p^{2})^{2}\,}\,.
\end{equation}

\subsection{General charge configuration}
Let us now consider the~$st^{2}$ AEs (\ref{st^2-AEs}) for a
generic BH charge configuration. By using the geometrical data
(\ref{st^2-geometry}), such Eqs. can be recast in the following
form:
\begin{equation}
\left\{
\begin{array}{l}
\partial _{s}V_{BH}=2\overline{\mathcal{Z}}\mathcal{D}_{s}\mathcal{Z}+iC_{stt}\left( g^{t\bar{t}}\right) ^{2}\left( \overline{\mathcal{D}}_{\bar{t}}\overline{\mathcal{Z}}\right) ^{2}=0;
\\[0.3em]
\partial _{t}V_{BH}=2\overline{\mathcal{Z}}\mathcal{D}_{t}\mathcal{Z}+2iC_{stt}g^{s\bar{s}}g^{t\bar{t}}\left( \overline{\mathcal{D}}_{\bar{s}}\overline{\mathcal{Z}}\right) \overline{\mathcal{D}}_{\bar{t}}\overline{\mathcal{Z}}=0.
\end{array}
\right. \label{st^2-AEs-2}
\end{equation}
Let us consider the non-BPS~$\mathcal{Z}=0$ class of attractors.
It is clear then that the only possibility for the BH effective
potential to have an extremum is the following one:
$$ \mathcal{Z}=0, \qquad \mathcal{D}_{t}\mathcal{Z}=0.$$
The corresponding AEs read
\begin{equation}
\left\{
\begin{array}{l}
\ds q_{0}+q_{1}s+q_{2}t+p^{0}st^{2}-p^{1}t^{2}-2p^{2}st=0, \\[0.5em]
\ds 2q_{0}+2q_{1}s+q_{2}\left( t+\bar{t}\right)
+2p^{0}st\bar{t}-2p^{1}t\bar{t}-2p^{2}s\left( t+\bar{t}\right) =0,
\end{array}
\right. \label{st^2-non-BPS-Z=0-AEs}
\end{equation}
which solutions are
\begin{eqnarray}
&& s=\frac{p^i q_i - 2p^{1}q_{1}\pm i \sqrt{\strut
\mathcal{J}_4}}{2\left[ \left( p^{2}\right) ^{2}-p^{0}q_{1}\right]
},
\qquad t=\frac{p^i q_i - 2 p^{2}q_{2}\mp i\sqrt{\strut \mathcal{J}_4}}{2p^{1}p^{2}-p^{0}q_{2}},\label{st^2-non-BPS-Z=0-solutions}\\
&& \mathcal{J}_4 =\mathcal{J}_4(p,q) \equiv
-(p^{0}q_{0}+p^{1}q_{1})^{2}+(2p^{1}p^{2}-p^{0}q_{2})(2p^{2}q_{0}+q_{1}q_{2}).\label{J4-st^2}
\end{eqnarray}
The newly appeared expression for~$\mathcal{J}_4 $ is nothing but
unique quartic invariant of the~$U$-duality group~$SU(1,1)^2$ in
the symplectic charge basis. The value of the entropy is given by
\be\label{entropy,st2} S=\pi \sqrt{\strut \mathcal{J}_4(p,q)}. \ee
Let us note that sign twist yields that the critical value of the
moduli~(\ref{st^2-non-BPS-Z=0-solutions}) with~$\mathcal{Z}=0$
relate to the corresponding ones with~$\mathcal{Z}\neq0$ by the
complex conjugation of some of them.

The BH charge configurations supporting the
non-BPS~$\mathcal{Z}=0$ attractors of the~$st^{2}$ model satisfy
the following constraints (for an analysis of the supporting BH
charge orbits see~\cite{BFGM1}):
\begin{equation}
\mathcal{J}_{4,st^{2}}\left( p,q\right) >0,~\left( p^{2}\right)
^{2}-p^{0}q_{1}\lessgtr 0,~2p^{1}p^{2}-p^{0}q_{2}\gtrless 0.
\label{st^2-non-BPS-Z=0-charge-constraints}
\end{equation}

\section{\label{stu}$stu$ Model}

In the so-called~$stu$ model the scalar manifold is the rank-$3$
homogeneous symmetric space~$\left(
\frac{SU(1,1)}{U(1)}\right)^{3}$ ($dim_{\Bbb{C}}=3$),
parameterized by the moduli~$z^{1}\equiv s$,~$z^{2}\equiv t$
and~$z^{3}\equiv u$ with negative imaginary parts~\cite{Gilmore}.
The corresponding $d=5$ parent theory is based on the rank-$2$
real special manifold $\left( SO\left( 1,1\right) \right) ^{2}$.
In special coordinates the prepotential reads $f=stu$.

The $stu$ model exhibits the noteworthy \textit{triality
symmetry}, in which all three moduli~$s$,~$t$ and~$u$ are on the
same footing and it has been studied in~\cite
{Duff-stu,BKRSW,K3,Ferrara-Marrani-1,TT2,Ferrara-Marrani-2}. Due
to this symmetry all expressions acquire quite elegant form.

It can be obtained as the element $n=0$ of the reducible cubic
sequence of homogeneous symmetric SK manifolds
$\frac{SU(1,1)}{U(1)}\otimes \frac{SO(2,2+n)}{SO(2)\otimes
SO(2+n)}$ (see \textit{e.g.}~\cite{BFGM1} and Refs. therein).
Furthermore, it is contained in all homogeneous (not necessarily
symmetric) SK $d$-geometries.

For a general BH charge configuration the superpotential, BH
effective potential, contravariant metric and $C$-tensor are
respectively given by:
\begin{equation}
\begin{array}{l}
\ds W(s,t,u)=q_{0}+q_{1}s+q_{2}t+q_{3}u+p^{0}stu-p^{1}tu-p^{2}su-p^{3}st\,, \\[0.5em]
\ds V_{BH}=-i\,\frac{|W(s,t,u)|^2 + |W(\bar
s,t,u)|^2 + |W(s,\bar t,u)|^2 +|W(s,t,\bar u)|^2}{\left( s-\bar{s}\right) \left( t-\bar{t}\right) \left( u-\bar{u}\right) }, \\
\ds
g^{i\bar{j}}=-\diag((s-\bar{s})^{2},(t-\bar{t})^{2},(u-\bar{u})^{2}),
\qquad C_{stu}=-\frac{i}{(s-\bar{s})(t-\bar{t})(u-\bar{u})}.
\end{array}
\label{stu-geometry}
\end{equation}

Correspondingly, the three complex AEs have the form
\begin{equation}\label{stu-AEs}
\begin{array}{l}
\ds W(\bar s, \bar t, u) W(\bar s, t, \bar u) - 2 W(\bar s, t, u) W(\bar s, \bar t, \bar u) = 0,\\[0.2em]
\ds W(\bar s, \bar t, u) W(s, \bar t, \bar u) - 2 W(s, \bar t, u) W(\bar s, \bar t, \bar u) = 0,\\[0.2em]
\ds W(s, \bar t,\bar u) W(\bar s, t, \bar u) - 2 W(s, t, \bar u)
W(\bar s, \bar t, \bar u) = 0
\end{array}
\end{equation}
reflecting the above mentioned \textit{triality symmetry} of the
model.

\subsection{Magnetic charge configuration}
Attractor equations~(\ref{stu-AEs}) are a system of nonlinear
equations of sixth order. Therefore their solving in a general
case constitutes a certain problem. Let us start by putting
$p^{0}=0$ and~$q_{1}=q_{2}=q_{3}=0$. All $\frac{1}{2}$-BPS and
some non-BPS~$\mathcal{Z}\neq 0$ solutions can be obtained using
\BPSA~\cite{TT,Shmakova,BKRSW}. Using a hint dropped by the~$st^2$
model, we will search only for solutions to the AEs~(\ref
{stu-AEs}) of the form:
\begin{equation}
z_{cr}^{i}\left( q,p\right) =\pm p^{i}\lambda \left( q,p\right).
\label{generalized-TT-Ansatz}
\end{equation}
with a choice of signs depending on~$i$. By choosing the same sign
for all~$i$, one recovers the \BPSA~(\ref{TT-Ansatz}). One should
stress that solutions of the form~(\ref{generalized-TT-Ansatz})
for the magnetic charge configuration were as well derived
in~\cite{K3}.

By triality symmetry, without loss of generality we can assume
\begin{equation}
s=p^{1}\lambda ,\qquad t=-p^{2}\lambda ,\qquad u=p^{3}\lambda .
\label{stu-generalized-TT-Ansatz}
\end{equation}
Consequently, by solving the $stu$ AEs (\ref{stu-AEs}) with
$p^{0}=0\,$, $q_{1}=q_{2}=q_{3}=0$ and assuming
(\ref{stu-generalized-TT-Ansatz}), one gets for $\lambda \left(
q_{0},p^{1}p^{2},p^{3}\right) $ the same solutions as above,
namely:
\begin{equation}\label{stu-non-Anz}
\begin{array}{lll}
1:& \quad\ds \left\{ \lambda _{1}=0,\quad \lambda _{2}=\pm \sqrt{\frac{q_{0}}{p^{1}p^{2}p^{3}}}\right\} ,&\quad q_{0}p^{1}p^{2}p^{3}>0, \\[0.5em]
2:&\quad\ds \left\{ \lambda _{1}=0,\quad \lambda _{2}=\pm
\sqrt{-\frac{q_{0}}{p^{1}p^{2}p^{3}}}\right\} ,&\quad
q_{0}p^{1}p^{2}p^{3}<0.
\end{array}
\end{equation}

Solution 1 is non-BPS $\mathcal{Z}=0$, and it is stable (no
Hessian massless
modes~\cite{Ferrara-Marrani-1,Ferrara-Marrani-2}).

Solution 2 is non-BPS $\mathcal{Z}\neq 0$, with
superpotential~$W=2q_{0}$. The corresponding Hessian matrix
splits~\cite{TT} in four massive and two massless modes and all
the consideration done above for the solution $2$ hold here, as
well.

For all solutions the minimal value of the BH effective potential
is
\begin{equation}
 V_{BH}\rule[-0.7em]{0.2pt}{1.6em}_{\,cr}=\frac1{\pi}S_{BH}=2\sqrt{\left| q_{0}p^{1}p^{2}p^{3}\right| \,}\,.
\end{equation}

\subsection{General Charge Configuration}
Let us now consider the $stu$ AEs (\ref{stu-AEs}) for a generic BH
charge configuration. By using the geometrical data
(\ref{stu-geometry}), such equations can be recast in the
following form:
\begin{equation}
\left\{
\begin{array}{l}
\ds \partial _{s}V_{BH} =
2\overline{\mathcal{Z}}\mathcal{D}_{s}\mathcal{Z}+iC_{stu}g^{t\bar{t}}g^{u\bar{u}}\left(
\overline{\mathcal{D}}_{\bar{t}}\overline{\mathcal{Z}}\right)
\overline{\mathcal{D}}_{\bar{u}}\overline{\mathcal{Z}}=0,
\\[0.4ex]
\ds \partial _{t}V_{BH} =
2\overline{\mathcal{Z}}\mathcal{D}_{t}\mathcal{Z}+iC_{stu}g^{s\bar{s}}g^{u\bar{u}}\left(
\overline{\mathcal{D}}_{\bar{s}}\overline{\mathcal{Z}}\right)
\overline{\mathcal{D}}_{\bar{u}}\overline{\mathcal{Z}}=0,
\\[0.4ex]
\partial _{u}V_{BH}=2\overline{\mathcal{Z}}\mathcal{D}_{u}\mathcal{Z}+iC_{stu}g^{s\bar{s}}g^{t\bar{t}}\left( \overline{\mathcal{D}}_{\bar{s}}\overline{\mathcal{Z}}\right) \overline{\mathcal{D}}_{\bar{t}}\overline{\mathcal{Z}}=0.
\end{array}
\right. \label{stu-AEs-2}
\end{equation}
Now it is clear that to find a zero central charge solution to
these equations it is enough to make equal to zero two out of
three components of the covariant derivative vector~${\mathcal
D}_i{\mathcal Z}$. Since the triality symmetry one can choose, for
example,
$${\mathcal D}_t{\mathcal Z}= {\mathcal D}_u{\mathcal Z}= 0.$$
Therefore the total set of AEs read as follows
\begin{equation}\label{stu-non-BPS-Z=0-AEs}
W(s,t,u) = W(s,\bar t, u ) = W(s,t,\bar u) = 0
\end{equation}
and has the following solution
\begin{equation}
s=\frac{p^i q_i - 2 p^{1}q_{1}\pm i \sqrt{\strut
\mathcal{J}_4}}{2\left( p^{2}p^{3}-p^{0}q_{1}\right) }, \quad
t=\frac{p^i q_i - 2p^{2}q_{2}\mp i\sqrt{\strut \mathcal{J}_4}
}{2\left( p^{1}p^{3}-p^{0}q_{2}\right) }, \quad \quad u=\frac{p^i
q_i - 2p^{3}q_{3}\mp i\sqrt{\strut \mathcal{J}_4} }{2\left(
p^{1}p^{2}-p^{0}q_{3}\right) }. \label{stu-non-BPS-Z=0-solutions}
\end{equation}
The expression for the entropy is the same as for the~$st^2$
model~(\ref{entropy,st2}), but with quartic invariant of the
$U$-duality group $\left( SU\left( 1,1\right) \right) ^{3}$ now
given by
\begin{equation}
\mathcal{J}_{4,stu}\left( p,q\right) \equiv -\left( p\cdot
q\right) ^{2}+4\left(
p^{1}q_{1}p^{2}q_{2}+p^{1}q_{1}p^{3}q_{3}+p^{2}q_{2}p^{3}q_{3}\right)
-4p^{0}q_{1}q_{2}q_{3}+4q_{0}p^{1}p^{2}p^{3}. \label{J4-stu}
\end{equation}
In~\cite{Duff-Cayley} this invariant was recognized to be nothing
but (the opposite of) the so-called Cayley's
hyperdeterminant~\cite {Cayley}. This fact leads to interesting
developments, relating the physics of extremal BHs to quantum
information theory~\cite
{KL
}-\cite{Ferrara-Duff3}.

As it occurs in the~$st^2$ model, for the~$stu$ model
non-BPS~$\mathcal{Z}=0$ solutions are related to those
with~$\mathcal{Z}\neq 0$ by the complex conjugations of some
moduli.

The BH charge configurations supporting the
non-BPS~$\mathcal{Z}=0$ attractors of the $stu$ model satisfy the
following constraints (for an analysis of the supporting BH charge
orbits see~\cite{BFGM1}):
\begin{equation}
\mathcal{J}_{4,stu}\left( p,q\right)
>0,~p^{2}p^{3}-p^{0}q_{1}\lessgtr 0,~p^{1}p^{3}-p^{0}q_{2}\gtrless
0,~p^{1}p^{2}-p^{0}q_{3}\gtrless 0.
\end{equation}

It should be noted that the~$stu$ AEs (\ref{stu-AEs-2}) can be
cast in the following form~\cite{FKlast}:
\begin{equation}
z_{1}z_{2}+{\bar z}_3{\bar z}_4=0, \quad z_{1} z_{3}+{\bar z_2}
{\bar z_4}=0, \quad z_{1}z_{4}+{\bar z_2}{\bar z_3}=0,
\label{stu-AEs-3}
\end{equation}
$$z_{1}\equiv -i\mathcal{Z}, \quad z_{2}\equiv -i\left( s-\bar{s}\right) \overline{\mathcal{D}_{s}\mathcal{Z}}, \quad
z_{3}\equiv -i\left( t-\bar{t}\right)
\overline{\mathcal{D}_{t}\mathcal{Z}}, \quad z_{4}\equiv -i\left(
u-\bar{u}\right) \overline{\mathcal{D}_{u}\mathcal{Z}} $$

The~$stu$ AEs in the form~(\ref{stu-AEs-3}) are identical to the
AEs of~$\mathcal{N}=8$,~$d=4$ supergravity, where $z_{i}$ are the
four complex eigenvalues of the central charge matrix~$Z_{AB}$
($A,B=1,\ldots,8$), which can be skew-diagonalized~\cite
{FKlast,BFGM1}. $\mathcal{N}=2$ $\frac{1}{2}$-BPS and
$\mathcal{N}=2$ non-BPS $\mathcal{Z}=0$ solutions, both having
$\mathcal{J}_{4} >0$, are lifted to
$\mathcal{N}=8$~$\frac{1}{8}$-BPS solutions, with positive quartic
Cartan-Cremmer-Julia invariant of the fundamental representation
$\mathbf{56} $ of the $\mathcal{N}=8$, $d=4$~$U$-duality group
$E_{7\left( 7\right) }$~\cite{Cartan, Cremmer:1979up}. On the
other hand, $\mathcal{N}=2$ non-BPS $\mathcal{Z}\neq 0$
solutions~$\mathcal{J}_4<0$, are lifted to $\mathcal{N}=8$ non-BPS
solutions, with, with the negative quartic Cartan-Cremmer-Julia
invariant. Such relations among $\mathcal{N}=2$
and~$\mathcal{N}=8$ attractors have been analyzed
in~\cite{ADF2,ADF3} for the BPS case and
in~\cite{Ferrara-Marrani-1} for the non-BPS case.

One should recall that the moduli~$s$,~$t$ and~$u$ can be
interchanged through triality symmetry; the explicit signs in the
relations~(\ref {stu-non-BPS-Z=0-solutions}) is due to the
``polarization'' introduced by picking
$\mathcal{D}_{s}\mathcal{Z}\neq 0$ in solving the non-BPS
$\mathcal{Z}=0$ AEs (\ref{stu-non-BPS-Z=0-AEs}). From an
$\mathcal{N}=8$ perspective, $\mathcal{N}=2$ non-BPS
$\mathcal{Z}=0$ attractors are originated from $\mathcal{N}=2$
$\frac{1}{2}$-BPS ones by simply exchanging two eigenvalues of the
skew-diagonal matrix $Z_{AB}$, namely $z_{1}$ and $z_{2}$ in the
conventions used above. Thus, the exchanging of two eigenvalues of
the $\mathcal{N}=8$ central charge matrix leads to complex
conjugation of the descendant attractor solutions in the
$\mathcal{N}=2$ supergravity obtained by performing a consistent
supersymmetry truncation $\mathcal{N}=8\longrightarrow
\mathcal{N}=2$.

\section{\label{Cayley-hyperdeterminant}Cayley's Hyperdeterminant and its
relation with $\mathcal{J}_{4,t^{3}}$
and~$\mathcal{J}_{4,st^{2}}$}

In this Section we reconsider Cayley's hyperdeterminant, and point
out its relation with the invariant $\mathcal{J}_{4,t^{3}}$ (of
the $U$-duality group $SU\left( 1,1\right) $) of the
so-called~$t^{3}$ model~\cite {Saraikin-Vafa-1} and with the
invariant $\mathcal{J}_{4,st^{2}}$ of the $st^{2}$ model, given by
Eq. (\ref{J4-st^2}).

Let us start by recalling once again that the $stu$ model is based
on the rank-$3$ homogeneous symmetric reducible SK manifold
$\frac{G_{4,stu}}{H_{4,stu}}=\left( \frac{SU(1,1)}{U(1)}\right)
^{3}$. It also holds that $SU(1,1)\sim SL\left( 2,\Bbb{R}\right)
\sim Sp\left( 2,\Bbb{R}\right) $, and that the symplectic group of
the $stu$ model is $Sp\left( 2n_{V}+2,\Bbb{R}\right) =Sp\left(
8,\Bbb{R}\right) $. The embedding of the~$U$-duality group
$G_{4,stu}=\left( SU(1,1)\right) ^{3}$ into~$Sp\left(
8,\Bbb{R}\right) $ is encoded by the spinor $\Psi _{\alpha \beta
\gamma }$, sitting in a real, spin $s=\frac{3}{2},$ $\left(
\frac{1}{2},\frac{1}{2},\frac{1}{2}\right) $-representation of
$G_{4,stu}$, with $\alpha $, $\beta $, $\gamma =0,1$. $\Psi
_{\alpha \beta \gamma }$ has no particular properties of symmetry
on its indices, thus it may get $2^{3}=8$ possible values,
\textit{i.e.} $8$ possible states corresponding to
the~$2n_{V}+2=8$ components of the BH charge vector in the
symplectic basis
\begin{equation}
Q_{stu}\equiv \left(
q_{0},q_{1},q_{2},q_{3},p^{0},p^{1},p^{2},p^{3}\right)
_{stu}=\left( q_{\Lambda },p^{\Lambda }\right) _{stu,\Lambda
=0,1,2,3}. \label{Q-sympl}
\end{equation}
Thus, $\left\{ \Psi _{\alpha \beta \gamma }\right\} _{\alpha
,\beta ,\gamma =0,1}$ is a basis for the BH charges, other than
the symplectic one (\ref {Q-sympl}). The relation between such two
basis is given in by Eq. (3.5) of~\cite{KL}.

The quartic (and unique) invariant of the $\left(
\frac{1}{2},\frac{1}{2},\frac{1}{2}\right) $-representation of
$G_{4,stu}$ can be defined as (see Eq. (2.8) of~\cite{KL})
\begin{equation}
\mathcal{J}_{4,stu}\equiv \frac{1}{2}\Psi _{\alpha _{1}\alpha
_{2}\alpha _{3}}\Psi _{\beta _{1}\beta _{2}\beta _{3}}\Psi
_{\gamma _{1}\gamma _{2}\gamma _{3}}\Psi _{\delta _{1}\delta
_{2}\delta _{3}}\epsilon ^{\alpha _{1}\beta _{1}}\epsilon ^{\alpha
_{2}\beta _{2}}\epsilon ^{\gamma _{1}\delta _{1}}\epsilon ^{\gamma
_{2}\delta _{2}}\epsilon ^{\alpha _{3}\gamma _{3}}\epsilon ^{\beta
_{3}\delta _{3}},
\end{equation}

where $\epsilon $ is the metric of $SU(1,1)$. In particular,
notice that no quadratic invariant of~$G_{4,stu}$ exists, because
$\Psi _{\alpha _{1}\alpha _{2}\alpha _{3}}\Psi _{\beta _{1}\beta
_{2}\beta _{3}}\epsilon ^{\alpha _{1}\beta _{1}}\epsilon ^{\alpha
_{2}\beta _{2}}\epsilon ^{\alpha _{3}\beta _{3}}=0$. In the
symplectic basis $\mathcal{J}_{4,stu}$ can be rewritten as given
by Eq. (\ref{J4-stu}). Thus,~$\mathcal{J}_{4,stu}$ can actually be
recognized~\cite{Duff-Cayley} as the opposite of the so-called
Cayley's hyperdeterminant~\cite{Cayley}:
\begin{equation}
\mathcal{J}_{4,stu}=-Det\left( \Psi \right) .
\end{equation}

A manifestly $SO(2,2)$-invariant form of $\mathcal{J}_{4,stu}$ can
be obtained by going to the so-called \textit{hatted} symplectic
basis:
\begin{equation}
\widehat{Q}_{stu}\equiv \left(
\widehat{q}_{0},\widehat{q}_{1},\widehat{q}_{2},\widehat{q}_{3},\widehat{p}^{0},\widehat{p}^{1},\widehat{p}^{2},\widehat{p}^{3}\right)
_{stu}=\left( \widehat{q}_{\Lambda },\widehat{p}^{\Lambda }\right)
_{stu,\Lambda =0,1,2,3},
\end{equation}

whose relation with the symplectic basis (\ref{Q-sympl}) is given
by Eq. (59) of~\cite{BKRSW}. Notice that $\widehat{q}_{\Lambda }$
and $\widehat{p}^{\Lambda }$ fit two distinct copies of the same
vector representation of $SO(2,2)$. In the $\widehat{Q}$-basis
$\mathcal{J}_{4,stu}$ reads (see Eq. (63) of~\cite{BKRSW})
\begin{equation}
\mathcal{J}_{4,stu}=\left( \widehat{p}\right) ^{2}\left(
\widehat{q}\right) ^{2}-\left( \widehat{p}\cdot \widehat{q}\right)
^{2},
\end{equation}

where $\left( \widehat{p}\right) ^{2}\equiv \left(
\widehat{p}^{0}\right) ^{2}+\left( \widehat{p}^{1}\right)
^{2}-\left( \widehat{p}^{2}\right) ^{2}-\left(
\widehat{p}^{3}\right) ^{2}$, and analogously for $\left(
\widehat{q}\right) ^{2}$.

Since the $stu$ model is the $n=0$ element of the cubic reducible
sequence $\frac{SU(1,1)}{U(1)}\otimes
\frac{SO(2,2+n)}{SO(2)\otimes SO(2n)}$ ($n_{V}=n+3$, $n\in
\Bbb{N}\cup \left\{ 0,-1\right\} $) of homogeneous symmetric SK
manifolds, it is easy find that the manifestly
$SO(2,2+n)$-invariant of the quartic (and unique) invariant of the
$U$-duality group $SU(1,1)\otimes SO(2,2+n)$ reads as follows:
\begin{equation}
\mathcal{J}_{4,SU(1,1)\otimes SO(2,2+n)}=\left( \widehat{p}\right)
_{n}^{2}\left( \widehat{q}\right) _{n}^{2}-\left( \widehat{p}\cdot
\widehat{q}\right) _{n}^{2},
\end{equation}

where the scalar products are now taken in $SO(2,2+n)$:
\begin{equation}
\begin{array}{l}
\left( \widehat{p}\cdot \widehat{q}\right) _{n}\equiv
\sum_{\Lambda =0,1}^{n+3}p^{\Lambda }q_{\Lambda }, \quad\left(
\widehat{p}\right) _{n}^{2}\equiv \left( \widehat{p}^{0}\right)
^{2}+\left( \widehat{p}^{1}\right) ^{2}-\left(
\widehat{p}^{2}\right) ^{2}-...-\left( \widehat{p}^{n+3}\right)
^{2},
\end{array}
\end{equation}
and analogously for $\left( \widehat{q}\right) _{n}^{2}$. In this
case, $\widehat{q}_{\Lambda }$ and $\widehat{p}^{\Lambda }$ fit
two distinct copies of the same vector representation of
$SO(2,2+n)$. Clearly,~$\left. \mathcal{J}_{4,SU(1,1)\otimes
SO(2,2+n)}\right| _{n=0}=\mathcal{J}_{4,stu}$.

It should also be recalled that $G_{4,stu}$ in general acts
linearly only in the basis $\left\{ \Psi _{\alpha \beta \gamma
}\right\} _{\alpha ,\beta ,\gamma =0,1}$, where each index is
related to a different factor $SU(1,1)$ inside $G_{4,stu}$.

Let us now consider the quartic (and unique) invariants of
the~$U$-duality groups of the $t^{3}$
model~\cite{Saraikin-Vafa-1}, based on the rank-$1$ homogeneous
symmetric SK manifold $\frac{SU(1,1)}{U(1)}$, and of the~$st^{2}$
model, based on the rank-$2$ homogeneous symmetric SK manifold
$\left( \frac{SU(1,1)}{U(1)}\right) ^{2}$.

In the symplectic basis
\begin{equation}
Q_{t^{3}}\equiv \left( q_{0},q_{1},p^{0},p^{1}\right)
_{t^{3}}=\left( q_{\Lambda },p^{\Lambda }\right) _{t^{3},\Lambda
=0,1},
\end{equation}

the quartic (and unique) invariant of the $U$-duality
group~$SU(1,1)$ of the $t^{3}$ model, denoted
by~$\mathcal{J}_{4,t^{3}}$, reads~\cite {Saraikin-Vafa-1}

\begin{equation}
\mathcal{J}_{4,t^{3}}\left( p,q\right) \equiv
-\frac{4}{9}p^{0}\left( q_{1}\right)
^{3}+\frac{1}{3}(p^{1})^{2}\left( q_{1}\right)
^{2}-(p^{0})^{2}\left( q_{0}\right)
^{2}-2p^{0}q_{0}p^{1}q_{1}+\frac{4}{3}(p^{1})^{3}q_{0}.
\end{equation}

On the other hand, in the symplectic basis
\begin{equation}
Q_{st^{2}}\equiv \left( q_{0},q_{1},q_{2},p^{0},p^{1},p^{2}\right)
_{st^{2}}=\left( q_{\Lambda },p^{\Lambda }\right) _{st^{2},\Lambda
=0,1,2},
\end{equation}

the quartic (and unique) invariant of the $U$-duality
group~$\left( SU(1,1)\right) ^{2}$ of the $st^{2}$ model, denoted
by $\mathcal{J}_{4,st^{2}}$, is given by Eq. (\ref{J4-st^2}).

The basis $\left\{ \Psi _{\alpha \beta \gamma }\right\} _{\alpha
,\beta ,\gamma =0,1}$ for the BH charges of the $stu$ model allows
one to relate $\mathcal{J}_{4,stu}$ with $\mathcal{J}_{4,t^{3}}$
and $\mathcal{J}_{4,st^{2}} $.

The relation between $\mathcal{J}_{4,stu}$
and~$\mathcal{J}_{4,t^{3}}$ can be simply achieved by totally
symmetrizing the $3$-index spinor $\Psi _{\alpha \beta \gamma }$:
\begin{equation}
stu:\Psi _{\alpha \beta \gamma }\longrightarrow t^{3}:\Psi
_{\left( \alpha \beta \gamma \right) },
\label{total-symmetrization}
\end{equation}
yielding $p^{1}=p^{2}=p^{3}$and $q_{1}=q_{2}=q_{3}$ in the $stu$
model. In such a way, the \textit{reduced} symplectic BH charge
vector of the $stu$ model becomes
\begin{equation}
\widetilde{Q}_{stu}\equiv \left. Q_{stu}\right|
_{p^{1}=p^{2}=p^{3},q_{1}=q_{2}=q_{3}}=\left(
q_{0},q_{1},p^{0},p^{1}\right) _{stu}
\end{equation}
In the (semi)classical limit of real, large BH charges, one
assumes in general $\widetilde{Q}_{stu}$ and~$Q_{t^{3}}$ to be
related by the following \textit{anisotropic rescaling}

\begin{equation}
\left( q_{0},q_{1},p^{0},p^{1}\right) _{stu}=\left( \eta
_{1}q_{0},\eta _{2}q_{1},\eta _{3}p^{0},\eta _{4}p^{1}\right)
_{t^{3}},~\eta _{1},\eta _{2},\eta _{3},\eta _{4}\in \Bbb{R}.
\end{equation}

By comparing $\left. \mathcal{J}_{4,stu}\right|
_{p^{1}=p^{2}=p^{3},q_{1}=q_{2}=q_{3}}$
with~$\mathcal{J}_{4,t^{3}}$, one obtains $\eta _{1}=\left( \eta
_{3}\right) ^{-1}$, $\eta _{2}=\left( 9\eta _{3}\right) ^{-1/3}$
and $\eta _{4}=\left( \eta _{3}/3\right) ^{1/3}$. Thus, in the
symplectic basis, $\mathcal{J}_{4,t^{3}}$ can be obtained from
$\mathcal{J}_{4,stu}$ by starting from $Q_{stu}$ and performing
the following identifications and rescalings ($\eta _{3}\in
\Bbb{R}_{0}$):
\begin{equation}
\left\{
\begin{array}{l}
q_{0,stu}=\left( \eta _{3}\right) ^{-1}q_{0,t^{3}}; \quad
q_{1,stu}=q_{2,stu}=q_{3,stu}=\left( 9\eta _{3}\right)
^{-1/3}q_{1,t^{3}};
\\
\\
p_{stu}^{0}=\eta _{3}p_{t^{3}}^{0}; \quad p_{stu}^{1}=\left( \eta
_{3}/3\right) ^{1/3}p_{t^{3}}^{1}.
\end{array}
\right. \label{stu-->t^3}
\end{equation}

On the other hand, the relation between $\mathcal{J}_{4,stu}$ and
$\mathcal{J}_{4,st^{2}}$ can be simply achieved by symmetrizing
the $3$-index spinor $\Psi _{\alpha \beta \gamma }$ only on two
indices, say (by triality symmetry, without loss of generality)
the first two ones:
\begin{equation}
stu:\Psi _{\alpha \beta \gamma }\longrightarrow st^{2}:\Psi
_{\left( \alpha \beta \right) \gamma },
\label{partial-symmetrization}
\end{equation}
yielding $p^{2}=p^{3}$and $q_{2}=q_{3}$ in the $stu$ model. In
such a way, the \textit{reduced} symplectic BH charge vector of
the $stu$ model becomes
\begin{equation}
\breve{Q}_{stu}\equiv \left. Q_{stu}\right|
_{p^{2}=p^{3},q_{2}=q_{3}}=\left(
q_{0},q_{1},q_{2},p^{0},p^{1},p^{2}\right) _{stu}
\end{equation}
As above, in the (semi)classical limit of real, large BH charges,
one assumes in general $\breve{Q}_{stu}$ and $Q_{st^{2}}$ to be
related by the following \textit{anisotropic rescaling}

\begin{equation}
\left( q_{0},q_{1},q_{2},p^{0},p^{1},p^{2}\right) _{stu}=\left(
\theta _{1}q_{0},\theta _{2}q_{1},\theta _{3}q_{2},\theta
_{4}p^{0},\theta _{5}p^{1},\theta _{6}p^{2}\right)
_{st^{2}},~\theta _{1},...,\theta _{6}\in \Bbb{R}.
\end{equation}

By comparing $\left. \mathcal{J}_{4,stu}\right|
_{p^{2}=p^{3},q_{2}=q_{3}}$ with $\mathcal{J}_{4,st^{2}}$, one
obtains $\theta _{1}=\left( \theta _{4}\right) ^{-1}$, $\theta
_{2}=\left( \theta _{5}\right) ^{-1}$, $\theta _{3,\pm }=\pm
\frac{1}{2}\left( \theta _{5}/\theta _{4}\right) ^{1/2}$ and
$\theta _{6,\pm }=\pm \left( \theta _{4}/\theta _{5}\right)
^{1/2}$. Thus, in the symplectic basis, $\mathcal{J}_{4,st^{2}}$
can be obtained from $\mathcal{J}_{4,stu}$ by starting from
$Q_{stu}$ and performing the following identifications and
rescalings ($\theta _{4},\theta _{5}\in \Bbb{R}_{0}$, $\theta
_{4}\theta _{5}>0$):
\begin{equation}
\left\{
\begin{array}{l}
q_{0,stu}=\left( \theta _{4}\right) ^{-1}q_{0,st^{2}}; \quad
q_{1,stu}=\left( \theta _{5}\right) ^{-1}q_{1,st^{2}}; \quad
q_{2,stu}=q_{3,stu}=\pm \frac{1}{2}\left( \theta _{5}/\theta
_{4}\right)
^{1/2}q_{2,st^{2}}; \\
\\
p_{stu}^{0}=\theta _{4}p_{st^{2}}^{0}; \quad p_{stu}^{1}=\theta
_{5}p_{st^{2}}^{1}; \quad p_{stu}^{2}=p_{stu}^{3}=\pm \left(
\theta _{4}/\theta _{5}\right) ^{1/2}p_{st^{2}}^{2}.
\end{array}
\right. \label{stu-->st^2}
\end{equation}

The positions (\ref{stu-->t^3}) and (\ref{stu-->st^2}) on
the~$stu$ BH charges are the simplest ones determining the
\textit{degeneracy} of the quartic invariant~$\mathcal{J}_{4,stu}$
(which is actually the opposite of the Cayley's hyperdeterminant
$Det\left( \Psi \right) $) into $\mathcal{J}_{4,t^{3}}$ and
$\mathcal{J}_{4,st^{2}}$, respectively. Such positions have a nice
interpretation as symmetrization of indices in the basis $\left\{
\Psi _{\alpha \beta \gamma }\right\} _{\alpha ,\beta ,\gamma
=0,1}$ for the $stu$ BH charges, because they respectively
correspond to Eqs. (\ref {total-symmetrization}) and
(\ref{partial-symmetrization}).

\section{\label{Conclusion}Conclusion}

In this work we derived the explicit non-BPS $\mathcal{Z}=0$
solutions to Attractor Equations for generic BH charge
configurations in two homogeneous symmetric $\mathcal{N}=2$, $d=4$
supergravity models, namely the so-called $st^{2}$ and $stu$ models.

Such non-supersymmetric solutions with vanishing central charge
were so far unknown, for two main reasons:
\begin{enumerate}
\item the non-BPS $\mathcal{Z}=0$ class of attractors is not present in the so-called $t^{3}$ model (so far, the only one in which the AEs
were explicitly and completely solved for generic BH charge
configurations~\cite {Saraikin-Vafa-1}),

\item the non-BPS $\mathcal{Z}=0$ class of attractors does not satisfy the so-called \BPSA. Indeed, it may be shown that the non-BPS $\mathcal{Z}=0$ constraints (\ref {non-BPS-Z=0-conds}) are not consistent with the
assumption (\ref{TT-Ansatz}) in the framework of a generic cubic
SK geometry.
\end{enumerate}

Furthermore, for the considered models we found that the non-BPS
$\mathcal{Z}=0$ critical moduli are (partially) related by complex
conjugation to their $\frac{1}{2}$-BPS counterparts, whose
explicit form for a generic SK $d$-geometry was known
since~\cite{BKRSW} and~\cite{Shmakova}. Moreover, such a relation
has a nice $\mathcal{N}=8$ interpretation in terms of exchange of
two of the four eigenvalues of the $\mathcal{N}=8$ central charge
matrix $Z_{AB}$.

It is worth pointing out that the non-BPS $\mathcal{Z}=0$
attractors of $\mathcal{N}=2$, $d=4$ ungauged supergravity might
have noteworthy phenomenological implications, because they
correspond to stabilized, purely charge-dependent configurations
of the scalars at the BH event horizon (with geometry
$AdS_{2}\times S^{2}$) for which the horizon maximal
$\mathcal{N}=2$ supersymmetry algebra~$\mathfrak{psu}(1,1\left|
2\right) $ is not centrally extended.

A possible field of application of the analytical formul\ae\ (\ref
{st^2-non-BPS-Z=0-solutions}) and
(\ref{stu-non-BPS-Z=0-solutions}) for non-BPS $\mathcal{Z}=0$
critical moduli in the $st^{2}$ and $stu$ models is the recently
established connection between the thermodynamics of extremal BHs
in supergravity and quantum information theory (also concerning,
for homogeneous symmetric supergravities, the so-called Jordan
algebras; see
\textit{e.g.}~\cite{BFGM1},~\cite{Rios},~\cite{Ferrara-Gimon} and
Refs. therein). As pointed out in Sect. \ref{stu}, such a relation
stems out from the observation, firstly made in~\cite
{Duff-Cayley}, that the quartic (and unique) invariant
$\mathcal{J}_{4,stu}\left( p,q\right) $ of the $U$-duality group
$\left( SU\left( 1,1\right) \right) ^{3}$ in the symplectic charge
basis is nothing but the Cayley's hyperdeterminant~\cite{Cayley}.
In this sense, the explicit relations (derived in
Sect.~\ref{Cayley-hyperdeterminant}) between Cayley's
hyperdeterminant and the quartic invariants of the $U$-duality
groups of the $t^{3}$~\cite{Saraikin-Vafa-1} and $st^{2}$ models
might be relevant for further developments. Interesting advances
along this line of
research have been achieved in the last months~\cite{KL%
}-\cite{Levay-SAM07}, and hopefully many others are to come in
next future.

\section*{\textbf{Acknowledgments}}

We would like to thank S. Ferrara for enlightening and fruitful
discussions.

The work of S.B., E.O and A.S. has been supported in part by the
European Community Human Potential Program under contract
MRTN-CT-2004-005104 \textit{``Constituents, fundamental forces and
symmetries of the universe''}.

The work of A.M. has been supported by a Junior Grant of the
\textit{``Enrico Fermi''} Center, Rome, in association with INFN
Frascati National Laboratories.


\end{document}